\documentstyle[12pt]{article}
\newcommand{\be}{\begin{equation}}
\newcommand{\ee}{\end{equation}}
\newcommand{\ba}{\begin{array}}
\newcommand{\ea}{\end{array}}
\newcommand{\bd}{\begin{displaymath}}
\newcommand{\ed}{\end{displaymath}}
\newcommand{\bt}{\begin{tabular}}
\newcommand{\et}{\end{tabular}}

\newcommand{\bc}{\begin{center}}
\newcommand{\ec}{\end{center}}

\begin{document}

\large
\bc
   {\bf     ON THE CALCULATION \\[3mm]
            OF THE VACUUM ENERGY DENSITY IN SIGMA MODELS   }
\ec
\vspace{10mm}

\bc
             V.G.KSENZOV
\ec
\vspace{5mm}

\bc
   {\it Institute of Theoretical and Experimental Physics,\\
        Moscow, Russia }
\ec
\vspace{12mm}

\bc
        A b s t r a c t
\ec
\vspace{8mm}

The vacuum energy density is calculated for the $O(N)$ nonlinear sigma models
in two dimensions. To obtain $\varepsilon_{vac}$ we assume that each
point of the space in which non-perturbative f\/ields are determined can be
replaced by a sphere $S^2$  having a small radius $r$ which approaches
zero at the very end of the calculation. This assumption allows to get
the classical f\/ields generating v.e.v. of the trace of the
energy-momentum tensor.
\vspace{20mm}

For a long time it has been clear that the critical phenomena in the
theory of condensed matter and the quantum f\/ield theory are related to the
conformal symmetry [1]. The symmetry breaking is def\/ined by means of the
trace of the energy-momentum tensor $\theta_{\mu \mu}$.
Using  $\theta_{\mu \mu}$
the phase structure of the $\sigma$-models in $2 \pm \varepsilon$
dimensions was investigated in [2]. The vacuum expectation value (v.e.v.) of
$\theta_{\mu \mu}$ gives the vacuum energy density $\varepsilon_{vac}$
and the characteristic mass scale  for theories without dimension parameters
[3].

$\varepsilon_{vac}$ may be def\/ined as
\bd
    \varepsilon_{vac} = \frac{1}{D}  ~<~ 0 ~| \theta_{\mu \mu}
    | ~0 ~> ~,
\ed
where $D$ is the dimension of the space.

To obtain v.e.v. of the $\theta_{\mu \mu}$ we have to know the role of
non-perturbative f\/luctuations in generating the physical amplitude. As a
rule, the inf\/luence of instantons on v.e.v. of dif\/ferent operators is
discussed because they are the sole non-perturbative f\/luctuations which
are known in QCD. However, there is a set of models, such as the $O(N)$
nonlinear sigma models, which do not have  instantons ($N > 3$),
but do have the  non-perturbative contributions [3, 4].

To understand the actual dynamics  and the role of non-perturbative ef\/fects
it is reasonable to study them on simple models.

For nonlinear sigma models some results were obtained in [5]
in case of large N. In this
paper it was shown that the non-perturbative contributions to the vacuum
energy density are connected with the square of the symmetry current of the
group $O(N)$. This result is valid for $\sigma$-f\/ields, which are subject
to the second class constraint.

The purpose of the present paper is to calculate $\varepsilon_{vac}$
in the framework of the two - dimensional $\sigma$-models
in the large $N$ limit. To do this we should have an explicit form
of the non-perturbative f\/ields.

The physical idea of the calculation is based on the critical assumption that
each point of the space, in which the non-perturbative f\/ields are determined,
can be replaced by a sphere $S^2$ having a small radius $r$ which
is set to zero at the very end of the calculation. This simple method allows
to build non-perturbative f\/ields in explicit form. From dimensional
considerations $\varepsilon_{vac}$ has to be of the order of $1/r^2$, therefore
naive $\varepsilon_{vac}$ goes to inf\/inity when $r$ tends to zero. Indeed,
the value $\varepsilon_{vac}$ is f\/inite due to some quantum f\/luctuations
which are orthogonal to classical f\/ields. It will be shown that the quantum
f\/ield correlator gives the value $exp(-4 \pi/f)$ and thus
we retrieve the familiar
result [3]. Then non-perturbative part of the vacuum energy density
$\varepsilon^n_{vac}$ is written as
\bd
   \varepsilon^n_{vac} = \frac{N}{8 \pi}  M^2 exp (-4 \pi / f) ~.
\ed

We recall that the perturbative part of the vacuum energy density \\
$\varepsilon^p_{vac}$  is def\/ined as
\bd
      \varepsilon^p_{vac} = - \frac{N}{8 \pi}  M^2 ~,
\ed
where $M^2$ is the regulator mass.

The Lagrangian of the $O(N)$ $\sigma$-models in 2 dimensions is taken
in the form
% 1
\be
   L = \frac{N}{2f}  \left(  \partial_{\mu} \sigma^a \right)^2 ,
   ~~~~~~~~~~~~~~~~~a = 1, 2, ... , N
\ee
and the $N$-component f\/ields are subject to the constraint
% 2
\be
   \sigma^a  \sigma^a = 1 ~.
\ee

The constraint introduces the interaction between f\/ields.

The action $S_E$ for  $\sigma$-models and the partition function
$Z_E [{\cal J}^a]$ in euclidian space are the following:
% 3
\be
\ba{c}
   Z_E [{\cal J}^a] =  \int D \sigma (x) D \alpha (x)
          exp (- S_E + \int d^2 x
          {\cal J}^a (x) \sigma^a (x) ) ~, \\ [4mm]
   S_E = \frac{1}{2} \int d^2 x \frac{N}{f} \left( \partial_{\mu}
         \sigma^a \right)^2 + \frac{\alpha (x)}{\sqrt N}
         \left(  \sigma^a \sigma^a - 1 \right) ~.
\ea
\ee

The factor $N^{-1/2}$ under the Lagrange multiplier f\/ield $\alpha (x)$
is written for the sake of convenience.

In order  to set of\/f the non-perturbative contributions in the vacuum
energy density one decomposes $\sigma^a(x)$ and $\alpha(x)$ as
$$
  \sigma^a = C^a (x) + q^a (x) ~,
$$
$$
  \alpha (x) = \alpha_C (x) + \alpha_{qu} (x)~,
$$
where $C^a (x)$ and $\alpha_C (x)$ are classical f\/ields, while $q^a (x)$
and $\alpha_{qu} (x)$ describe small f\/luctuations around the classical
background. Under this decomposition in the large $N$ limit, we get
% 4
$$
   Z_E = exp \left( -S_{cl} + \int d^2 x {\cal J}^a (x) C^a (x) \right)
   \int D q^a (x) D \alpha_{qu} (x)
$$
\be
   exp \left( -S_{qu} + \int
   d^2 x {\cal J}^a (x) q^a (x) \right) ~.
\ee
Here,
$$
  S_{cl} = \frac{1}{2} \frac{N}{f} \int d^2 x \left( \partial_{\mu} C^a
       \right)^2 ~,
$$
$$
  S_{qu} = \frac{1}{2} \int d^2 x  \frac{N}{f} \left( \partial_{\mu}
   q^a \right)^2  + \frac{\alpha_C (x) q^a q^a}{\sqrt N} +
   \frac{\alpha_{qu}}{\sqrt N} \left( C^a q^a
    + q^a q^a \right) ~.
$$
The equations of motion for the $c^a (x)$ and $\alpha_C (x)$
f\/ields are
% 5
\be
\ba{rcl}
   \frac{N}{f} \partial^2 C^a & = & \frac{\alpha_C C^a}{\sqrt N}
       - {\cal J}^a ~,        \\ [3mm]
   C^a C^a & = & 1
\ea
\ee
and for the quantum f\/ields
% 6
\be
\ba{l}
    2 C^a q^a + q^a q^a = 0 ~, \\[4mm]
    \frac{N}{f} \partial^2 q^a = \frac{\alpha_C q^a}{\sqrt N}
    - {\cal J}^a + \frac{\alpha_{qu}}{\sqrt N} \left( C^a + q^a \right) ~.
\ea
\ee
Let us simplify  eqs.(6) using them in the linear form with respect to the
quantum f\/luctuations. We have
% 7
\be
\ba{c}
   C^a q^a = 0 ~, \\ [4mm]
   \frac{N}{f} \partial^2 q^a = \frac{\alpha_C}{\sqrt N} q^a +
    \frac{\alpha_{qu} C^a}{\sqrt N} ~.
\ea
\ee
We are interested in v.e.v. of $\theta_{\mu \mu}$, therefore ${\cal J}^a$
is taken  equal to zero. Besides, the integral over $\alpha_{qu}$ in eq.(4)
can be replaced by $\delta (C^a q^a)$, that is we are integrating over $q^a$
being subject to the constraint (7).

It is known that v.e.v. of $\theta_{\mu \mu}$ can be obtained by
varying $ln Z_E$
according to the change of the coupling constant on $\delta f = \xi \beta (f)$,
where $\xi$ - is the parameter of the global scale transformation and
$\beta (f) = -M^2 \frac{df}{dM^2}$  [3, 6]
% 8
\be
\ba{c}
     < 0 | \theta_{\mu \mu}  | 0 >
      =  \frac{\beta (f)}{2 f^2}  N ~< 0 | \left(
     \partial_{\mu} q^a \right)^2 | 0 > ~,        \\[4mm]
     < 0 | \left( \partial_{\mu} q^a \right)^2 | 0 > ~=~
     Z_E^{-1} exp (-S_{cl})  \int D q^a \delta (c^a q^a) (\partial_{\mu}
     q^a)^2  exp (-S_{qu}) ~.
\ea
\ee
Here,
$S_{qu} = \frac{1}{2} \int d^2 x  \frac{N}{f} (\partial_{\mu} q^a)^2
+ \frac{\alpha_C}{\sqrt N} q^a q^a$.\\
Notice that we omit the term which is obtained by varying the classical
action. As we shall see later, it will give the contribution to the
pertiturbative part of the vacuum energy.

The problem of the calculation of the quantum correlation $< (\partial_{\mu}
q^a)^2 >$ is to take into account the condition $C^a q^a = 0$ and to do
a regularization of the quantum correlation.

The condition of orthogonality of quantum and classical f\/ields
can be satisf\/ied  by using the identity
% 9
\be
    \partial_{\mu} C^a = - J^{ab}_{\mu} C^b ~.
\ee
Here, $ J^{ab}_{\mu} = C^a \partial_{\mu} C^b - \partial_{\mu} C^a C^b$.
This is true when the f\/ields are subject to the constraint $C^aC^a = 1$.
Dif\/ferentiating $C^a q^a = 0$ and using (9) we obtain
% 10
\be
   C^a \left(  \partial_{\mu} q^a +  J^{ab}_{\mu} q^b \right) ~=~ 0 ~.
\ee

For arbitrary $C^a$, eq.(10) is satisf\/ied if
% 11
\be
   \partial_{\mu} q^a + J^{ab}_{\mu} q^b  ~=~ 0~~~~~
    {\mbox or}
\ee
$$
  q^a (x) = P exp \left(  - \int\limits^x_0 d z^{\mu} \hat{J}_{\mu}
  \right)_{ab}  q^b (0) ~.
$$
Hence,  $ q^a (x) q^a (x) = q^a (0) q^a (0)$.

Then we can obtain
$$
  < 0 | \left(  \partial_{\mu} q^a \right)^2 | 0 > ~=~
  J^{ak}_{\mu} J_{\mu}^{ap} < 0 | q^k (x)
  q^p (x) | 0 >
$$
and with due regard for
$$
  < 0 | q^k (x) q^p (x) | 0 > ~=~  \delta^{pk} < 0 | q^i (x)
   q^i (x) | 0 > ~,
$$
we get
% 12
\be
   < 0 | \left( \partial_{\mu} q^a \right)^2 | 0 > ~=~
   2 \left( \partial_{\mu} C^a \right)^2 < 0 | q^i (x) q^i (x) | 0 > ~.
\ee
Here use was made of the identity
$\left(  J^{ab}_{\mu} (x) \right)^2 =
2 \left( \partial_{\mu} C^a (x) \right)^2$.
There is no sum on the repeated indices. We also suppose that for any
indices $i = 1, 2, ..., N$ all correlators $< 0 | q^{i} (x) q^{i} (x) | 0 >$
are the same.

{}From eqs.(8) and (12) we get
% 13
\be
   <  0 | \theta_{\mu \mu} | 0 > = \frac{\beta (f)}{f^2} N
    \left( \partial_{\mu} C^a (x) \right)^2  < 0 | q^{i} (x)
    q^{i} (x) | 0 > ~.
\ee

In eq.(13) the magnitude of $<  0 | \theta_{\mu \mu} | 0 >$  is constant,
at the same time the quantity $\left( \partial_{\mu} C^a (x) \right)^2$
is a function of $x$ in a general case. In such case we have to find
particular classical f\/ields $C^a (x)$ which satisfy the condition
$\left( \partial_{\mu} C^a \right)^2 = const$ and the constraint (5).

The f\/irst condition may be written in the form
% 14
\be
   \partial^2 C^a (x) ~=~ const C^a (x) ~.
\ee
To f\/ind some non-trivial solution to eq.(14) with account of
 (5) let us consider the
special case when $N = 3$ and each point $x = x_0$ in the two-dimensional
euclidean space, in which eq.(14) is def\/ined, is replaced by a sphere
$S^2$ with a small radius $r$ in eucludean 3 dimensional
space and center in $x_0$.

The quantity $r$ is a new parameter which is set to zero at the very
end of the calculation. From dimension considerations, eq.(14)
must  be written as
% 15
\be
    \partial^2 C^a ~=~ \frac{2 C^a}{r^2 } ~.
\ee

It can be verif\/ied that the solution of eq.(15) is the function
{}~$C^a = \frac{z^a}{r}, \\ z^a = (x - x_0)^a,~~ a = 1, 2, 3,~~ r^2 = (z^a)^2$.
 So the solution coincides with the asymptotics of the
Higgs f\/ields for the monopoles of 't Hooft and Polyakov [7, 8].
The asymptotics of vector f\/ields for the monopoles
also coincides with the symmetry
current in our models. In addition, the identity (9), in this case,
coincides with the condition of the parallel transport of the Higgs
f\/ields along way $dx$, i.e. with the asymptotics of the covariant derivative.
It may well be monopole of 't Hooft-Polyakov is def\/ined
in a sphere $S^2$ which has asymptotics at the distance $r$.
It is not surprising because introducing of sphere $S^2$ in euclidean space
is associated with the emergence of a non-trivial second homotopy
group $\pi_2 (O (3)) = Z_2$.

To obtain a solution of eq.(15) for any $N > 3$ we can choose the matrix of
$C$-f\/ields in such a way that the f\/irst three f\/ields $C^i (i = 1, 2, 3)$
coincide with the solution for $O(3)$-group and the rest of $N$-
f\/ields is some constants.
These $C$-f\/ields have to be subject to the constraint (5) and
therefore the f\/ields $C^i$  have to be normalized as $C^i C^i = g$.
The multitude of the other special solutions can be obtained by rotating
the solution in global $O(N)$ group.

We have obtained that the term $(\partial_{\mu} C^a)^2$ is of the order of
$\mu^2 = 1/r^2$. Therefore the varying of the classical action
determines the perturbative part of the vacuum energy.

To calculate the magnitude $< 0 | q^i (x_0) q^i (x_0) | 0 >$ we have
to regularize it by separating the quantum f\/ields in dif\/ferent points.
That is, instead of the magnitude $< 0 | q^i (x_0) q^i (x_0) | 0 >$
we have to calculate the value
$$
   \lim_{\Delta \to 0}
   < 0 | q^i (x_0 - \Delta) q^i (x_0 + \Delta) | 0 > ~.
$$
Using eq.(11) we obtain

$$
  \lim_{\Delta \to 0}  < 0 | q^i (x_0 - \Delta) q^i (x_0 + \Delta) | 0 >
   =  < 0 | q^k (0) q^k (0) | 0 > \lim_{\Delta \to 0}
   \left( - \int\limits^{x+ \Delta}_{x-\Delta} d x_{\mu} \hat J_{\mu}
   \right)_{ik} =
$$
$$
   =  < 0 | q^k (0) q^k (0) | 0 >  exp
   \left( - 2 \oint d x_{\mu} \hat J_{\nu}  \right)_{ik} ~.
$$
The coefficient 2 emerges at the circled integral because there are two
different types of the circles. One of them is passed clockwise and the
other is passed in the opposite direction. We sum over this types
of the circles.
The circled integral is $\oint d x_{\mu} \hat J_{\nu} = \oint d C^K C^i
- d C^i C^K = 2 S_{ik} $. The area $S_{ik}$ is enclosed  by the circle
which is oriented normally to the third axis in the iso-space
and equals $\pi g$.

Let us denote the constant $< 0 | q^i (0) q^i (0) | 0 >$ by $\gamma$.
We arrive at

% 16
\be
   < 0 | \theta_{\mu \mu} | 0 > ~=~ g \frac{\beta (f)}{f^2}
   N \frac{1}{r^2} \gamma e^{-4 \pi g } ~.
\ee
Here, our result is defined on the scaling mass $\mu^2 = 1/r^2$.
The constant $\gamma$
is proportional to $\hbar$ and therefore $\varepsilon_{vac}$
approaches zero in the classical limit $(\hbar \to 0)$ as it must be for the
quantum conformal anomaly.

In order to obtain the net result we ought to obtain $\beta (f)$ and
to get the v.e.v. of $\theta_{\mu\mu}$ corresponding to $r^2 = 0$.
Notice that the classical fields $C^i$ defined on the scaling mass $\mu^2$
can be written in terms of the bare classical fields $C^i_0$ as
$C^i = Z^{1/2} C^i_0$. In this case the relation between $g$
and unrenormalized constant $g_0$ is written as $g= Zg_0$.
We also define relation $< 0 | \theta_{\mu \mu} | 0 > = Z < | \theta
_{\mu\mu} | 0 >_0$.

In order to find $Z$ we must get renormalized action $S_r$.
In the large $N$ limit the leading graph renormalizing the classical
action is the tadpole diagram in which we integrate over momentum
$p^2 \geq \mu^2$.

Thus we get
%17
\be
   S_r = \int d^2 x \frac{N}{f} (\partial_{\mu} C^i_0)^2
   (1 - \frac{f}{4 \pi} \ln \frac{M^2}{\mu^2})~ .
\ee
We recall that only the first three of $C$-fields are function of $x$.
Therefore $(\partial_{\mu} C^a)^2 = (\partial_{\mu} C^i)^2,
i = 1, 2, 3; a = 1, 2, ..., N. $

{}From eq.(17) we find
%18
\be
   Z =  1 - \frac{f}{4 \pi} \ln \frac{M^2}{\mu^2}~.
\ee
There is minimum of the action under condition $\mu^2 = m^2$,
where $m^2$  satisfies the equation
%19
\be
    \frac{f}{4 \pi}ln \frac{M^2}{m^2} ~=~ 1~.
\ee
Taking into consideration this equation we obtain
$$
   \frac{\beta (f)}{f^2} ~=~ \frac{1}{4 \pi}~ .
$$
Now, we get the final result
%20
\be
   < 0 | \theta_{\mu \mu} | 0 >_0 = \gamma g_0 \frac{N}{4 \pi}
   \mu^2  exp \left( - 4 \pi g_0 \left( 1 - \frac{f}{4 \pi}
    \ln \frac{M^2}{\mu^2} \right) \right)~.
\ee
We can see that this quantity is scaling independent only if $f g_0 = 1$.
This means that the quantity $g_0$ is not a free parameter. When $\mu^2 = M^2$,
we obtain
%21
\be
   \varepsilon_{vac} = \gamma g_0 \frac{N}{8\pi} M^2
   exp ( - 4 \pi / f)~.
\ee
The quantity  $\varepsilon_{vac}$ differs from the familiar result
[3] by the coefficient $\gamma g_0$. Therefore we may suppose that there is
only part of $\varepsilon_{vac}$ in eq.(21) which is defined by the
monopole of 't-Hooft-Polyakov.

We have studied the contribution of the non-perturbative f\/luctuation in
the vacuum energy density. It was shown that some classical f\/ields were
obtained provided that each point of the euclidean space has been replaced by
sphere $S^2$ with a small radius. This method was associated with the
emergence of the monopole in each such sphere because of the
existence of the non-trivial map $S^2 \to S^2$.

Due to the quantum f\/ields which are normal to the non-perturbative
f\/ields $\varepsilon^n_{vac}$ remains f\/inite when the sphere radius
tends to zero.
\vspace{5mm}

The author is grateful to V.N.Novikov and Yu.M.Makeenko for useful
discussions.
\vspace{5mm}

This paper was partially supported by the Russian Foundation of
Fundamental Research (grant N$^0$ 95-02-04681-A).

 \end{document}